\documentstyle[12pt]{article}
\normalbaselines

\begin{document}

\baselineskip=24pt

\bibliographystyle{unsrt}
\vbox {\vspace{6mm}}
\bigskip
\bigskip
\bigskip
\bigskip
\begin{center} {\bf PONDEROMOTIVE CONTROL OF\\ 
QUANTUM MACROSCOPIC COHERENCE}
\end{center}

\begin{center} 
{\it S. Mancini, V. I. Man'ko\footnote{On leave from Lebedev 
Physical Institute, Moscow}
 and P.Tombesi }
\end{center}

\begin{center}
Dipartimento di Matematica e Fisica, Universit\`a di Camerino, 
I-62032 Camerino\\
and\\
Istituto Nazionale di Fisica della Materia
\end{center}

\bigskip

\begin{center}
14 November 1996
\end{center}

\bigskip
\bigskip
\bigskip
\bigskip

\begin{abstract}
It is shown that because of the radiation pressure a 
Schr\"odinger cat state can be generated in a
resonator  with oscillating wall. The optomechanical
control of quantum macroscopic coherence  and its  
detection is taken into account introducing new
cat states. The effects due to the environmental couplings 
with this nonlinear system 
are considered developing an operator perturbation 
procedure to solve the 
master equation for the field mode density operator. 
\end{abstract}

PACS number(s): 42.50.Dv, 42.50.Vk, 03.65.Bz

\bigskip
\bigskip
\bigskip
\bigskip
\bigskip

\section{Introduction}\label{s1}

\noindent

One of the fundamental aspects of quantum mechanics 
is the existence of 
interference among quantum states which signs the 
difference between a 
superposition of states and a mixture of states. 
The quantum theory may 
adequately well describe macroscopic objects by 
means of a linear 
superposition of states with macroscopically distinguishable 
properties.
Recently, due to the improved technology, there was a growing 
interest on the possibility of
observing  such superposition states, commonly known as 
Schr\"odinger cats       
\cite{Schrodinger}. Good candidates for these macroscopic 
states are the 
coherent states of an e.m. field mode. The properties of 
superposition of two
generic coherent states has been studied in Ref. \cite{cahgla} 
and the simplest
superposition of even and odd coherent states was introduced in
Ref. \cite{dm74}. A review of these states is given in Ref. 
\cite{buz}. Within the
field of optics several proposals for the generation of linear 
superpositions 
of coherent states in various nonlinear processes 
\cite{YS,nonlinearcats} and 
in quantum non-demolition measurements \cite{QNDcats} have 
been made. It is
worthy to note that the field in a cat state has a lot of 
advantages in
optical communication \cite{hirota}.       
However, by coupling the system to its enviroment, as in 
the act of measurement, one always
introduces  dissipation and decoherence effects, which 
tend to destroy any quantum 
features \cite{CaldeiraLeggett}.

In the common scheme of the Kerr-like medium modeled by 
an anharmonic 
oscillator, it was shown \cite{DanielMilburn}  that the 
photon number 
distribution 
and interferences in phase space are highly sensitive to 
even small 
dissipative coupling. This fact, plus the smallness of
the $\chi^{(3)}$ nonlinearity, makes the prospect of 
experimentally producing 
and detecting such states highly questionable in these media. 

On the other hand, it is well known \cite{Meystre} that 
an empty optical 
cavity with a moving mirror may  mimic a Kerr-like medium 
when it is 
illuminated with coherent light. The effect of intensity 
dependent optical 
path is due, in this case, to the radiation pressure force. 

In this paper we shall present such a model as an 
alternative one for the 
generation of Schr\"odinger cats. We will show that, 
with the appropriate 
measurement technique, 
it could also be useful 
for revealing quantum macroscopic coherence.   

\section{The Model}\label{s2}

\noindent

We consider a linear Fabry-Perot empty cavity with one 
fixed partially 
reflecting end mirror and
one perfectly reflecting mirror, which can move 
(undergoing harmonic 
oscillations) under the
influence of radiation pressure. If $L$ is the 
equilibrium cavity length, 
the resonant 
frequency
of the cavity will be 
\begin{equation}\label{omc}
\omega_c=\pi\frac{c}{L}n
\end{equation}
where $n$ is an integer number determined by the 
frequency of the input 
light and $c$ is the speed of light.
We assume that the retardation effects, due to the 
oscillating mirror, in 
the intracavity field
are negligible. We will also neglect the correction 
to the radiation 
pressure force due to the
Doppler frequency shift of the photons \cite{Unruh}. 
Thus we are able to 
write the Hamiltonian of
the whole system as 
\begin{equation}\label{H}
H=\hbar\omega_c a^{\dag}a+\hbar\omega_m b^{\dag}b+H_{int}
\end{equation}
where $a,\,a^{\dag}$are the boson operators of the 
resonant cavity mode 
and $b,\,b^{\dag}$ are
the boson operators of the oscillating mirror with 
the mass $m$ and the 
angular frequency
$\omega_m$. This latter will be many order of magnitude 
smaller than 
$\omega_c$ to ensure that
the number of photons generated by the nonstationary
Casimir effect \cite{law} as consequence of the Casimir 
forces \cite{Casimir} in the resonator with moving boundaries  
is completely negligible.
$H_{int}$ accounts for the fact that the intracavity photon 
changes 
its energy, by $\omega_c$, as
the oscillating mirror moves \cite{ManTom}
\begin{equation}\label{Hint}
H_{int}= -\hbar G a^{\dag}a(b+b^{\dag})
\end{equation}
with the coupling constant given by
\begin{equation}\label{G}
G=\frac{\omega_c}{L}\left(\frac{\hbar}{2m\omega_m}\right)^{1/2}.
\end{equation}
From the Hamiltonians (\ref{H}) and (\ref{Hint}) we can derive, 
using the 
BCH formula for the Lie algebra \cite{Wilcox}, the time evolution 
operator in 
the following form
\begin{equation}\label{U}
U(t)=e^{iE(t)(a^{\dag}a)^2}e^{iF(t)
a^{\dag}a{\hat x}(t)}\left[e^{-i\omega_c a^{\dag}at/\omega_m}
e^{-ib^{\dag}bt}\right]
\end{equation}
where
\begin{equation}\label{x}
\hat x(t)=be^{it/2}+b^{\dag}e^{-it/2}
\end{equation}
is the mirror quadrature operator, while
\begin{equation}\label{EF}
E(t)=\kappa^2[t-\sin t]\,;\quad F(t)=2\kappa\sin(t/2)\,;\quad 
\kappa =G/\omega_m\,,
\end{equation}
with $t$ the time scaled by $\omega_m$, i.e. we have replaced 
$\omega_mt$ by $t$.
Furthermore, from now on, we will consider the evolution operator 
omitting the free motion of the
two modes $a$ and $b$, i.e. the term inside the square brackets 
on Eq. (\ref{U}).

\section{Generation of Schr\"odinger Cat States}\label{s4}

\noindent

From Eq. (\ref{U}) one can immediately recognizes that the time 
evolution 
introduces anharmonicity
due to the presence of the nonlinear term $(a^{\dag}a)^2$ whose 
strenght 
depends also on time
\cite{MTKerr}. It is also easy to see that at each time for which
$F(t)=0$ the two subsystems are disentagled. Furthermore due to its 
macroscopicity we should  consider
the oscillating mirror initially in a thermal state at 
temperature $T$
\begin{equation}\label{thermal}
\rho_T=(1-z)\sum_n z^n|n\rangle\langle n|;\quad
z=\exp\left(-\frac{\hbar\omega_m} {k_BT}\right)\,,
\end{equation}
with $z/(1-z)=N_{th}$ that represents the mean number of 
excitations of the mechanical oscillator,
i.e. the number of thermal phonons. Thus starting from an initial
coherent state 
$|\alpha_0\rangle$ for the radiation mode we have 
\begin{equation}\label{psi}
\rho(t^*)=e^{iE(t^*)(a^{\dag}a)^2}|\alpha_0\rangle|\langle\alpha_0|
\otimes\rho_T
e^{-iE(t^*)(a^{\dag}a)^2}
\end{equation}
with 
\begin{equation}\label{t*}
t^*=2\pi m_1;\quad m_1\in N
\end{equation}
so that 
\begin{equation}\label{EFvalues}
F(t^*)=0;\quad E(t^*)=\kappa^2 2\pi m_1.
\end{equation}
Now in order to see the cat states, the following condition must be 
fulfilled \cite{YS}
\begin{equation}\label{YScondition}
E(t^*)=\frac{\pi}{2}+2\pi m_2;\quad m_2\in N
\end{equation}
so that combining Eqs.(\ref{EFvalues}) and (\ref{YScondition}) 
one gets
\begin{equation}\label{k}
\kappa^2=\frac{1}{m_1}(\frac{1}{4}+m_2)
\end{equation}
which can be read as a restriction on the possible values of 
the various 
external parameters.
Thus if the above conditions are satisfied, we have
\begin{equation}\label{Psitstar}
\rho(t^*)=\frac{1}{2}\left[e^{-i\pi/4}|\alpha_0\rangle+
e^{i\pi/4}|-\alpha_0\rangle\right]
\left[\langle -\alpha_0| e^{-i\pi/4}+
\langle\alpha_0| e^{i\pi/4}\right]
\otimes \rho_T\,;
\end{equation}
however, this is not the only way to create a quantum 
superposition in this 
system. In fact, let us
consider the times $t'$ for which
\begin{equation}\label{Et'}
E(t')=\frac{\pi}{2}+2\pi m;\quad m\in N\,.
\end{equation}
In these cases, obviously, $F(t')$ is not necessarily zero then, the 
reconstruction of the
superposed coherent states is impossible due to the entanglement 
between 
the two subsystems.
One can now use a conditional measurement to create the desired 
states, 
performing a sort of
quantum state engineering \cite{Sch}.

Let us suppose that the mirror's quadrature $\hat x(t)$ is measured 
\cite{CavesThorne},
giving the result
$y_t$. The state of the radiation field after the measurement 
is found by 
projecting the
system's state onto the eigenstate $|y_t\rangle$
\begin{equation}\label{condsta}
\rho_{after}(t)={\cal C}e^{iE(t)(a^{\dag}a)^2}
e^{iF(t)a^{\dag}ay_t}|\alpha_0\rangle
\langle y_t|\rho_T|y_t\rangle\langle\alpha_0|
e^{-iF(t)a^{\dag}ay_t}|\
e^{-iE(t)(a^{\dag}a)^2}\,,
\end{equation}
where $\cal C$ is a normalization constant
\begin{equation}\label{N}
{\cal C}=(\langle y_t|\rho_T|y_t\rangle)^{-1}\,.
\end{equation}
At the times $t'$ we have, from Eqs. (\ref{Et'}) and
(\ref{condsta})
\begin{eqnarray}\label{Psit'}
\rho_{after}(t')=&\frac{1}{2}&\left[e^{-i\pi/4}
|\alpha_0e^{iF(t')y_{t'}}\rangle
+e^{i\pi/4}|-\alpha_0e^{iF(t')y_{t'}}\rangle\right]\nonumber\\
&\otimes&\left[\langle -\alpha_0e^{iF(t')y_{t'}}| e^{-i\pi/4}
+\langle\alpha_0e^{iF(t')y_{t'}}| e^{i\pi/4}\right]\,,
\end{eqnarray}
which is a superposition of coherent states whose phase depends 
on the measurement process; 
and further, if the result of the measurement is
\begin{equation}
y_{t'}=\frac{\pi}{2}\frac{1}{F(t')}\,,
\end{equation}
it is possible to recover in Eq. (\ref{Psit'}) the generalized 
even and odd
coherent states like those
discussed in \cite{Spiridonov,nieto} which show quantum 
interference as 
well as other particular features. 

\section{Quasiprobability and Marginal Distribution}

\noindent

The evolved density operator of the whole system can be easily 
constructed by using the 
time evolution operator of eq. (\ref{U})
\begin{equation}\label{rhoevov}
\rho(t)=U(t)|\alpha_0\rangle
\langle\alpha_0|\otimes\rho_TU^{\dag}(t)\,,
\end{equation}
and then the evolution can be described for example, in terms of 
the $Q$-function   
\begin{eqnarray}\label{Q}
&&Q(\alpha,\beta,t)=\langle\alpha|\langle\beta|\rho(t)
|\beta\rangle|\alpha\rangle=
e^{-|\alpha|^2-|\alpha_0|^2-|\beta|^2} (1-z)
\sum_{j=0}^{\infty}z^j|\beta|^j
\nonumber\\
&\times&\Bigg|\sum_{n=0}^{\infty}\frac{(\alpha^*\alpha_0)^n}{n!}
\exp\left\{\left[iE(t)-\frac{1}{2}F^2(t)\right]n^2+iF(t)ne^{-it/2}
\beta^*\right\}\sum_{r=0}^{j}\frac{\left(iF(t)ne^{it/2}\right)^r}
{r!\sqrt{(j-r)!}}\Bigg|^2\,,\nonumber\\
\end{eqnarray}
where the variables $\alpha$, $\beta$ are referred  to the 
radiation and to the mirror respectively.
However, since the distinguishing element of a linear 
superposition of 
coherent states is the
presence of interference fringes in the marginal distribution, 
we are 
interested in that one, for
the particular times discussed in the previous Section.
Its definition, for a generic state $\rho^{field} (t)$ of the
 radiation field, is given by
\begin{equation}\label{defmarg}
P(X)=\langle X|\rho^{field} (t)|X\rangle\,,
\end{equation}
where $|X\rangle$ are eigenstates of the quadrature operator 
$X=(a+a^{\dag})/2$, while
$\rho^{field}$ should be intended as ${\rm Tr}_m\{\rho\}$ with 
${\rm Tr}_m$ the trace over 
the mirror degrees of freedom.
In the case of Eq. (\ref{Psitstar}) we can integrate over the 
degree of 
freedom of the mirror to
obtain the marginal distribution of the field mode as 
\cite{DanielMilburn} 
\begin{eqnarray}\label{P}
P(X)&=&\left|\langle X|\frac{1}{\sqrt{2}}\left[e^{-i\pi/4}
|\alpha_0\rangle+
e^{i\pi/4}|-\alpha_0\rangle\right]\right|^2\nonumber\\
&=&\frac{1}{2}\left[P_+(X)+P_-(X)
+2\sqrt{P_+(X)P_-(X)}\sin\Big(4X|\alpha_0|\sin\left(
\arg\alpha_0\right)\Big)\right],\nonumber\\
\\
P_{\pm}(X)&=&\left(\frac{2}{\pi}\right)^{1/2}
\exp\left[-2X-|\alpha_0|^2\mp
2X(\alpha_0+\alpha_0^*)-\frac{1}{2}(\alpha_0^2
+\alpha_0^{*2})\right]\,;
\nonumber
\end{eqnarray}
while in the case of Eq. (\ref{Psit'}) the marginal 
distribution for the 
field mode is in effect
a conditional probability
\begin{equation}\label{Pcond}
P(X|y_{t'})=\left|\langle X|\frac{1}{\sqrt{2}}\left[e^{-i\pi/4}
|\alpha_0e^{iF(t')y_{t'}}\rangle+
e^{i\pi/4}|-\alpha_0e^{iF(t')y_{t'}}\rangle\right]\right|^2\,,
\end{equation}
whose explicit expression is the same as in Eq. (\ref{P}), 
apart an extra 
phase factor in
the coherent sate which gives the interference pattern along 
a direction 
depending on the result of the measurement as well.

\section{Damped Mode Equation and Solutions}

\noindent

Let us now consider the proposed model as an open system 
interacting with 
the "rest of Universe" \cite{Garbook}.
We will study only the case in which
the radiation mode relaxes much faster than the mirror
(the opposite case, i.e. the mirror 
that relaxes much faster
than the cavity mode, does not show any quantum features 
due to the 
thermalization effects).
Moreover,  since, in
order to see the Schr\"odinger cats, we are interested to 
short time 
behaviour (i.e. times much shorter than the typical radiation 
relaxation time), 
we can consider
the mirror practically  not
affected by any damping. Hence,
the master equation for the whole system will be taken in the 
form 
\begin{equation}\label{rhogen}
\dot\rho=\frac{i}{\hbar}[\rho,H]+\chi(\rho)\,,
\end{equation}
where \cite{CollGar}
\begin{equation}\label{chi}
\chi(\rho)=\frac{\gamma}{2}[2a\rho a^{\dag}-a^{\dag}a\rho
-\rho a^{\dag}a]\,,
\end{equation}
and where we have considered the number of thermal photons to be 
negligible at optical
frequencies. In our model, the damping constant $\gamma$ 
takes into account the loss of photons through the fixed mirror, 
so it is related to its
transmissivity  $Tr$ by the relation $\gamma=c Tr /2L$ with $c$ 
the speed of light. However, since
we are using a scaled time we should replace $\gamma/\omega_m
\rightarrow\gamma$.
Now, the undamped system is an exact solvable system with the free 
evolution
operator
$U(t)$ given by Eq. (\ref{U}) and obeying  the equation
$i\dot U(t)=HU(t)$.
Then,
introducing a new density operator $R$, in a form similar to 
the interaction 
representation, i.e.
$\rho=URU^{\dag}$, we may rewrite Eq. (\ref{rhogen}) as
\begin{equation}\label{Req}
\dot R=U^{\dag}\chi(URU^{\dag})U=\tilde\chi(R)\,,
\end{equation}
where the operator $\tilde\chi(R)$ is obtained by the following 
recepie: all the 
additional operators $a_i$
in the initial operators $\chi(\rho)$ are replaced by 
$\tilde a_i=U^{\dag}a_iU$,
while the operator $\rho$ is replaced by $R$.
We could write down the solution of the Eq. (\ref{Req}) in the form 
$R=R_0+Y$,
where $R_0$ is a constant operator, i.e. $\dot R_0=0$, and the 
operator $Y$
satisfies the equation corresponding to (\ref{Req})
\begin{equation}\label{Yeq}
\dot Y=\tilde\chi(Y+R_0)\,.
\end{equation}
The operator $R_0$ represents to the free solution of the initial 
Eq. (\ref{rhogen}), i.e.
without the term $\chi(\rho)$. Till now we only rewrote the 
master equation in
another representation and it is still an exact equation. However, 
Eq. (\ref{Req}) is appropriate
to apply the Born iteration procedure \cite{Merz} provided that the 
the damping term $\chi(\rho)$ is small enough to be considered 
as a perturbative one 
(this could be the case since the parameter $\gamma$ has to be
small in order to achieve the  Schr\"odinger cats).
Then we could try to solve Eq. (\ref{Yeq}) simply by replacing in 
the r.h.s. the operator
$\tilde\chi(Y+R_0)$ by
$\tilde\chi(R_0)$, i.e.  performing the first Born approximation. 
The solution is
immediate, and the operator $R$ assumes the form 
\begin{equation}\label{Rsol}
R(t)=R_0+\int_0^t\tilde\chi(R_0,\tau)d\tau\,.
\end{equation}
It means that the initial density operator $\rho(t)$ becomes
\begin{equation}\label{rhot}
\rho(t)=\rho_0(t)+\rho_{\gamma}(t)\,,
\end{equation}
where the term $\rho_0(t)$ is the density operator of the 
free motion 
\begin{equation}\label{rho0}
\rho_0(t)=U(t)\rho_0(0)U^{\dag}(t)
\end{equation}
with initial density matrix $\rho_0(0)\equiv\rho(0)$. 
The correction term  
$\rho_{\gamma}(t)$ has the form
\begin{equation}\label{rhog}
\rho_{\gamma}(t)=U(t)\left[\int_0^t\tilde\chi(R_0,\tau)d\tau\right]
U^{\dag}(t)\,,
\end{equation}
or more explicitely
\begin{eqnarray}\label{rhoga}
\rho_{\gamma}(t)&=&\gamma\int_0^t d\tau\left\{e^{-iF(t-\tau)
\hat x(t-\tau)-2iE(t-\tau)a^{\dag}a}
a\rho_0(t)a^{\dag}
e^{iF(t-\tau)\hat x(t-\tau)+2iE(t-\tau)a^{\dag}a}
\right\}\nonumber\\
&-&\frac{\gamma}{2}t\left[a^{\dag}a\rho_0(t)+
\rho_0(t)a^{\dag}a\right]\,.
\end{eqnarray}
The range of validity of the above approximation is determined 
by the 
requierement $\rho_{\gamma}(t)<<\rho_0(t)$.
Below, it will become more clear that it works for 
$\gamma|\alpha_0|^2t<<1$.
It is also easy to check that $\hbox{Tr}\{\rho_{\gamma}\}=0$, 
then
$\rho(t)$ is always normalized to
 unity. Let us now try to find the
marginal distribution at the particular
times $t^*$ and $t'$ discussed in Sec. \ref{s4}. By means 
of Eqs. (\ref{rhoga}),
(\ref{defmarg}) and (\ref{Psitstar}), after lenghtly but 
straithforward 
algebra, one obtains
\begin{eqnarray}\label{rhogat*}
&&P(X)=\langle X|{\rm Tr}_{m}\{\rho_0(t^*)
+\rho_{\gamma}(t^*)\}|X\rangle\nonumber\\
&&=\left(\frac{2}{\pi}\right)^{\frac{1}{2}}e^{-|\alpha_0|^2-2X^2}
\sum_{p,q=0}^{\infty}\frac{2^{-(p+q)/2}}{p!q!}
H_p(\sqrt{2}X)H_q(\sqrt{2}X)
e^{i{\rm arg}\alpha_0(p-q)}\nonumber\\
&&\times|\alpha_0|^{p+q}\left\{\frac{A_{q,p}}{2}+
\frac{\gamma}{2}
\left[A_{p,q}I_{p,q}(t^*)|\alpha_0|^2-A_{q,p}
\frac{p+q}{2}t^*\right]\right\}
\end{eqnarray}
where $H_p$ are the
Hermite polynomials,
\begin{equation}\label{Apq}
A_{p,q}=\left[1+i(-)^q-i(-)^p+(-)^{p+q}\right]\,,
\end{equation}
and finally
\begin{equation}\label{Ipq}
I_{p,q}(t^*)=\int_0^{t^*} d\tau e^{-i[2E(t^*-\tau)](p-q)}\,.
\end{equation}
In Eq. (\ref{rhogat*}) the first term inside the curly brackets 
comes from $\rho_0$ and is related
to the undamped motion, while the other is the perturbative term 
due to the environmental
coupling. Due to the fact that at the times
$t^*$ the two subsystems (i.e. radiation cavity mode and mirror) 
are 
disentangled, the thermal
effects do not destroy the cat state as can be seen in the 
above equations. 
The decoherence depends
only on the leakage of photons through the fixed mirror.
 
In the case of cat states generated by conditional measurement 
the expression for 
the conditional
probability in presence of damping has almost the same structure of 
Eq. (\ref{rhogat*}), and can be
obtained by using Eqs. (\ref{rhoga}), (\ref{defmarg}) and 
(\ref{Psit'})
\begin{eqnarray}\label{rhogat'}
&&P(X|y_{t'})=\langle X|\langle y_{t'}|\rho_0(t')
+\rho_{\gamma}(t')|y_{t'}\rangle|X\rangle
\nonumber\\
&&={\cal C'}\left(\frac{2}{\pi}\right)^{1/2}e^{-|\alpha_0|^2-2X^2}
\sum_{p,q=0}^{\infty}\frac{2^{-(p+q)/2}}{p!q!}
H_p(\sqrt{2}X)H_q(\sqrt{2}X)
e^{i[{\rm arg}\alpha_0+F(t')y_{t'}](p-q)}\nonumber\\
&&\times |\alpha_0|^{p+q}\left\{\frac{A_{q,p}}{2}\langle y_{t'}
|\rho_T|y_{t'}\rangle+
\frac{\gamma}{2}
\left[A_{p,q}\tilde I_{p,q}(t')|\alpha_0|^2-
A_{q,p}\frac{p+q}{2}t'\langle y_{t'}|\rho_T|y_{t'}\rangle 
\right]\right\}
\nonumber\\
\end{eqnarray}
where
\begin{eqnarray}\label{Itilde}
\tilde I_{p,q}(t')=\int_0^{t'}&d\tau &
\exp\{-i[2E(t'-\tau)+F(t')F(t'-\tau)\sin(\tau/2)](p-q)\}
\nonumber\\
&\times &
\langle y_{t'}-F(t'-\tau)\sin(\tau /2)|\rho_T|y_{t'}
-F(t'-\tau)\sin(\tau /2)\rangle\,,
\end{eqnarray}
and, due to Eq. (\ref{thermal}), the following general 
expression holds \cite{Morse}
\begin{equation}\label{ypsi}
\langle{\cal Y}|\rho_T|{\cal Y}\rangle=
\left(\frac{2}{\pi}\right)^{\frac{1}{2}}
(1-z)\sum_{j=0}^{\infty}\frac{z^j}{2^jj!}
e^{-2{\cal Y}^2}H^2_j(\sqrt{2}{\cal Y})=
\left(\frac{2}{\pi}\frac{1-z}{1+z}\right)^{\frac{1}{2}}
\exp\left[-2{\cal Y}^2\frac{1-z}{1+z}\right]\,.
\end{equation}
$\cal C'$ is a constant needed for the normalization after 
the projection in the measurement
process, and it can be obtained by performing 
the integration over the X variable of Eq. (\ref{rhogat'})
with the
aid of the completness formula for the Hermite polynomials 
\cite{Morse}
\begin{equation}\label{Pytilde}
{\cal C'}=\left\{\langle y_{t'}|\rho_T|y_{t'}\rangle +
\gamma |\alpha_0|^2\left[\tilde I_{p,q=p}(t')
-\langle y_{t'}|\rho_T|y_{t'}\rangle t'\right]
\right\}^{-1}\,.
\end{equation}
It is easy to note that the correction term in both 
solutions (\ref{rhogat*})
and (\ref{rhogat'}) remains smaller than the undamped 
term provided 
$\gamma|\alpha_0|^2t<<1$.
Equation (\ref{rhogat'}) shows a dependence of the
decoherence effects also on the
thermal state of the mirror (i.e. its temperature).
In Figs. (1) and (2), we show respectively $P(X)$ and 
$P(X|y_{t'}=0)$ (solid
lines) of Eqs. (\ref{rhogat*}) and (\ref{rhogat'})  
contrasted with the same in absence of damping
(dashed lines).  We may see that in the
case of the cat state created at $t^*=2\pi$, i.e. Fig. (1), 
the coherence 
has been almost totally
washed out, due to the long time needed for the formation; 
while the 
conditional measurement
could be used to generate the superposition at shorter time, 
in Fig. (2) $t'=3
\pi/2$, preserving the coherence effects. In this case, 
however, one should pay attention to the
thermal effect of the mirror. To this end, let us consider 
more closely the case of $y_{t'}=0$,
which is a high probable value for the mirror quadrature 
measurement. The normalization factor on
the r.h.s. of Eq. (\ref{ypsi}) is a common factor that 
can be eliminated in Eq. (\ref{rhogat'}) by
using Eq. (\ref{Pytilde}), while the exponentioal factor 
remains in the integral of Eq.
(\ref{Itilde}) only. As $z$ approaches the value 1, i.e. 
the temperature increases, it tends to
becomes unity. This means that the thermal effects tend 
to destroy the coherence only up to a value
of temperature, above which the interference fringes 
become insensitive (dotted line of Fig. (2)).
Of course analogous discussions can be made for other 
values of the mirror quadrature $y_{t'}$. 

We also note from both Fig. (1) and Fig. (2) that, as 
the dissipation becomes relevant, two gaussian
peaks centered around the mean number of photons, and 
which are typical of the orthogonal
quadrature, appear.
This is essentially due to the
rotation in the phase space introduced by the damping 
term $\rho_{\gamma}$.
In fact, as can be seen in Eq. (\ref{rhoga}), it involves 
an integration over the time which leads
to a distribution whose contributions come from various 
field phases, i.e. from different
quadratures. 

\section{Detection of Quantum Coherence}

\noindent

In this section we will show that the above discussed 
model could also be used 
to reveal the quantum coherence. 

According to Ref. \cite{Haroche}, the photon number 
statistics of the 
radiation field could be opportunely used as signature 
of the presence of 
Schr\"odinger cat states.
On the other hand, in the presented model, a
measurement of the mirror's momentum $\hat p$ allows us 
to get the photon number 
statistics in an indirect way \cite{JTWC}.
In particular the signal could be represented by the 
number $a^{\dag}a$ of photons
of the radiation mode, and the meter by the
momentum of the movable mirror; the out of phase 
quadrature coupled to the 
photon
number (see Eq. (\ref{Hint})).

Our purpose should be to detect the Schr\"odinger 
cat immediately after 
its generation
inside the cavity, at time
$t^*$ (or
$t'$ if one uses conditional measurement generation); 
nevertheless in both 
cases the two
subsystems, i.e. the mirror and the radiation mode, are 
disentangled
(as can be seen in Eqs. (\ref{Psitstar}) and (\ref{Psit'})), 
so no information
can be extracted in indirect way. 
Then, we must address the measurement to get something 
which is slightly 
different from
the Schr\"odinger cat state, but still having quantum 
coherence features. To 
this end, let us
consider at first the entanglement between the signal 
and the meter, 
which could be
described by the correlation function defined as 
follow \cite{Hol}
\begin{equation}\label{corrfunc}
C_{s,m}=\frac{|\langle a^{\dag}a\hat p\rangle
-\langle a^{\dag}a\rangle\langle \hat p\rangle|^2}
{V_{a^{\dag}a}V_{\hat p}}\,,
\end{equation}
where $V$ means the variance. This quantity shows 
how good is the scheme as a 
measurement
device, and should be equal to one for a perfect scheme. 
By performing 
the expetaction
values using Eqs. (\ref{rhoga}) and (\ref{U}) we obtain 
\begin{equation}\label{Csm}
C_{s,m}=
\frac{2|\alpha_0|^2\kappa^2\left[\sin^2t
+4\gamma\sin^2\left(\frac{t}{2}
\right)\sin t\right]}{\left[\frac{1}{2}+N_{th}+
2|\alpha_0|^2\kappa^2\sin^2t\right](1-\gamma t)
+|\alpha_0|^2\kappa^2 
\frac{\gamma}{2}\left[2t-8\sin t
+3\sin(2t)\right]}\,.
\end{equation}
Thus $C_{s,m}$ is a function of $t$ depending also 
on $\kappa$, which is a 
constant that contains all the external parameters.
Fig. (3) illustrates the typical behaviour of $C_{s,m}$ 
versus $t$,
showing the effects of dissipation as well as the thermal ones.
From this figure it is obviously that higher values of $C_{s,m}$ 
for times closer to $0,\pi,2\pi$
could be achieved by increasing the value of $\kappa$ or  
$|\alpha_0|$, but we must take into account that the number of 
photons plays a
delicate role in the dissipation effect.

Let us now consider a time at
which the radiation is entangled with the mirror, then its state, 
in absence of loss, by Eq.
(\ref{rhoevov}), will be 
\begin{eqnarray}\label{rhofield}
\rho_0^{field}(t)&=&{\rm Tr}_m\{\rho_0(t)\}\nonumber\\
&=&\int dy_t \langle y_t|\rho_T|y_t\rangle e^{iE(t)(a^{\dag}a)^2}
|\alpha_0e^{iF(t)y_t}\rangle\langle\alpha_0e^{iF(t)y_t}|
e^{-iE(t)(a^{\dag}a)^2}\,,
\end{eqnarray}
and furthermore if $E(t)$ satisfies the condition 
(\ref{YScondition}) for 
that time, it becomes
\begin{eqnarray}\label{rhopseudo}
&\rho_0^{field}(t)&
=\frac{1}{2}\int dy_t\langle y_t|\rho_T|y_t\rangle\nonumber\\
&\times&\left(e^{-i\frac{\pi}{4}}|\alpha_0e^{iF(t)y_t}\rangle
+e^{i\frac{\pi}{4}}|-\alpha_0e^{iF(t)y_t}\rangle\right)
\left(e^{-i\frac{\pi}{4}}\langle-\alpha_0e^{iF(t)y_t}|
+e^{i\frac{\pi}{4}}\langle\alpha_0e^{iF(t)y_t}|\right)\nonumber\\
\end{eqnarray}
which represents not a "pure" cat state, but that one whose 
phase is still 
convoluted with
the mirror motion and at which we may refer as "pseudo-cat" state. 
This latter, however, has the advantage of being detected, since 
it does not 
imply any
disentanglement.  It is worth to remark that
the dephasing effect due to the factor $\exp(iFy_t)$, which 
degrades the 
pure cat into
a pseudo-cat state, is considerable only for those values of $y_t$ 
contained under 
the gaussian
state of the mirror. 
Then the temperature can emphasize this 
negative effect, since
it introduces highest mirror number states, i.e. gaussians 
with larger 
width. On the other hand, in order to reduce this effect, 
it is also
preferable to have the 
smallest possible
values of $F(t)$. 
These are accessible only at times near to $2\pi$ 
(see eq. (\ref{EF})). 
Thus, in order to realize the measurement, the choice of 
the mesurement 
time $t$ and the
value of $\kappa$ should be made to fulfill simultaneously 
the following 
requirements:
Eq. (\ref{YScondition}), the highest value of $C_{s,m}$, and the 
smallest value of $F$. 
Of course the detection should be performed at time much 
shorter than the typical cavity
lifetime $\gamma^{-1}=2L/cTr$, but also longer than the 
photon cavity fly time $2L/c$, to ensure
the presence of photons inside the cavity. 

Let us now suppose to have found the desired $t$ and $\kappa$, 
then we 
revise the measurement strategy of Ref. \cite{Haroche} for 
the detection 
of quantum macroscopic coherence. 

A coherent field $|\alpha_r\rangle$, the "reference", is 
added to the 
pseudo-cat state,
immediately before the measurement, so that the resulting field in 
the cavity at the time
of measurement is
\begin{equation}\label{rhopseudoref}
\tilde\rho^{field}(t)
=\frac{1}{{\cal N}}D(\alpha_r)\rho^{field}D^{-1}(\alpha_r)\,,
\end{equation}
where $D$ is the displacement operator and
$\cal N$ is a normalization constant. 

After the injection of the reference field, the photon number 
distribution in the cavity becomes
\begin{eqnarray}\label{Pn}
&&{\cal P}(n)=\langle n|\tilde\rho^{field}(t)|n\rangle
=\frac{1}{{\cal N}}\frac{1}{2}\int dy_t\langle
y_t|\rho_T|y_t\rangle\nonumber\\ &&\times\Bigg\{
\left|e^{-i\frac{\pi}{4}}\langle n|\alpha_0e^{iF(t)y_t}
+\alpha_r\rangle
+e^{i\frac{\pi}{4}}\langle
n|-\alpha_0e^{iF(t)y_t}+\alpha_r\rangle\right|^2\Bigg\}
+O(\gamma)\,,
\end{eqnarray}
where $O(\gamma)$ indicates the perturbative terms 
proportional to the first power of $\gamma$
that we have omitted for space reasons. 

Let us now consider separately two cases. When
$\alpha_0$ and
$\alpha_r$ have  the same phase
the photon distribution, denoted by ${\cal P}_{in}(n)$, 
as consequence of the first term in Eq.
(\ref{Pn}), which is the dominant one, should appear as 
the sum  of two quasi
Poissonian distributions peaked around $n=|\alpha_0+\alpha_r|^2$ 
and
$n=|-\alpha_0+\alpha_r|^2$, with the tails due to the smearing 
effect of 
the gaussian
integral. In fact, in Eq. (\ref{Pn}), the interference part 
will be negligible provide
to have
$|\alpha_r|>>1$. An interesting situation arises when
$\alpha_0$ and $\alpha_r$ have the same amplitude, then
\begin{eqnarray}\label{Pin}
&&{\cal P}_{in}(n)=\frac{1}{{\cal N}}\frac{1}{2}
\frac{|\alpha_0|^{2n}}{n!}\int dy_t\langle y_t|\rho_T|y_t\rangle
\nonumber\\
&&\times\Bigg\{\left[c_+^ne^{-|\alpha_0|^2c_+}
+c_-^ne^{-|\alpha_0|^2c_-}
+2c_+^{n/2}c_-^{n/2}e^{-2|\alpha_0|^2}\Re\{-i(i)^n\}\right]\Bigg\}
+O(\gamma)\,,
\end{eqnarray}
where 
\begin{equation}
c_{\pm}=2\pm 2\cos\left(F(t)y_t\right)\,.
\end{equation}
In that case, neglecting the perturbation terms,
${\cal P}_{in}(n)$ consists of a very sharp distribution centered 
at $n=0$, which is a 
$\delta$-like peak
for a pure cat state, and a distribution peaked around 
$n=4|\alpha_0|^2$.
The existence of two separate peaks in the in-phase sum 
field is the proof 
of the
existence of two classical fields within the cavity.
However, it does not prove that these two fields are in a 
coherent quantum
mechanical superposition. So we need to consider also the case 
when $\alpha_0$ and
$\alpha_r$ are $\pi/2$ out of phase, for which we have
\begin{eqnarray}\label{Pout}
&&{\cal P}_{out}(n)=\frac{1}{{\cal N}}\frac{1}{2}
\frac{|\alpha_0|^{2n}}{n!}\int dy_t\langle y_t|\rho_T|y_t\rangle
\nonumber\\
&&\times\Bigg\{\left[s_+^ne^{-|\alpha_0|^2s_+}
+s_-^ne^{-|\alpha_0|^2s_-}
+2s_+^{n/2}s_-^{n/2}e^{-2|\alpha_0|^2}\Re\{-i(-i)^n\}
\right]\Bigg\}+O(\gamma)
\end{eqnarray}
where now 
\begin{equation}
s_{\pm}=2\pm 2\sin\left(F(t)y_t\right)\,.
\end{equation}
In this case the interference in the term in Eq.
(\ref{Pn}) becomes important, in fact ${\cal P}_{out}(n)$, 
again neglecting the perturbation terms,
exhibits a  Poisson envelope
with strong oscillations, signaling the coherence effect. 
The above 
discussed dephasing
effect in the pseudo-cat tends to wash out the oscillations 
and to 
transform the Poisson
envelope in a gaussian one. Of course in both cases 
($in$ and $out$) also the damping terms cause  a
degradation of the signal.

In Fig. (4) we show ${\cal P}_{in}(n)$ for a pseudo-cat state a) 
which resembles that one for a pure
cat state, contrasted with the same in presence of damping at 
zero temperature b) and at finite
temperature c).
Fig. (5) illustrates the same situations for ${\cal P}_{out}(n)$. 
Both figures are obtained using
$t=0.84\times 2\pi$ and $\kappa=0.5$ for which one has $F=0.48$ 
and $C_{s,m}=0.85$ (at zero
temperature, while it is reduced to 0.55 when $N_{th}=2$).

The ${\cal P}_{in}(n)$ and ${\cal P}_{out}(n)$ distributions 
can actually be
measured detecting the momentum of the mirror, of course the 
measurement 
process is
destructive, hence the state has to be reprepared for each 
measurement, 
and a
large number of measurements should be performed to reach 
the desired 
statistics.  
Then, from these output distributions, one can recognize a 
signature of quantum coherence as in
Fig. (4) and Fig. (5), provided to have small dissipation 
and very low temperature, which is needed
also to guarantee a sufficient signal meter correlation 
(Fig. (3)).

Finally, to effectively visualize the presence of interference 
fringes in the phase
space, we would consider the marginal distribution for the 
pseudo-cat. This probability, obtained through the expectation 
value $\langle X|
{\rm Tr}_m\{\rho_0(t)+\rho_{\gamma}(t)\}|X\rangle$ and using 
Eqs.(\ref{rhopseudo}) and 
(\ref{rhoga}), will be
\begin{eqnarray}\label{Ppseudo}
&&P^{pc}(X)=\left(\frac{2}{\pi}\right)^{1/2}
e^{-|\alpha_0|^2-2X^2}
\sum_{p,q=0}^{\infty}\frac{2^{-(p+q)/2}}{p!q!}
H_p(\sqrt{2}X)H_q(\sqrt{2}X)e^{i{\rm
arg}\alpha_0(p-q)}
\nonumber\\
&&\times|\alpha_0|^{p+q}\left\{\frac{A_{q,p}}{2}+
\frac{\gamma}{2}\left[A_{p,q}I_{p,q}(t)|\alpha_0|^2-
A_{q,p}\frac{p+q}{2}t\right]\right\}
\exp\left[-\frac{F^2(t)(p-q)^2(1+z)}{8(1-z)}\right],\nonumber\\
\end{eqnarray}
where, with the superscript $pc$ we refer to the pseudo-cat state.
It is clear from the last exponential factor how the thermal 
phonons of the mirror tend to
rapidly destroy the coherence effect.

In Fig. (6) we show the marginal distribution 
$P^{pc}(X)$ of Eq. (\ref{Ppseudo}) for various situations, using the 
above discussed values of
parameters i.e. $t=0.84\times 2\pi$, $\kappa=0.5$. 
From this picture we may note  that the interference pattern of the
pseudo-cat state is almost the same of the pure one
and is still
preserved at the time of measurement, even in the presence of loss 
provided to have a very small
number of thermal excitations in the mechanical oscillator.

\section{Conclusion}

\noindent

We have proposed the use of an optomechanical model for 
the generation of 
optical Schr\"odinger cat states. We have also presented a new 
scheme to 
reveal the quantum
macroscopic coherence, based on the new states named pseudo-cats 
that could be intended as 
a sort of cat states which could be recognized before their 
"natural birth".
Thus the model is substantially able to produce and also to 
detect interference effects without
introducing different couplings, but one should pay attention 
to the different sources of
dissipation.

We would also point
out that the studied system could be implemented for example  
idealizing the movable mirror as 
a piezoelectric crystal \cite{Fabre}.  The above
used values of $t$, $\kappa$ and $\gamma$ (in the various figures), 
could be reached for example with the
following set of parameters 
$\omega_c\approx 10^{16}\,{\rm s}^{-1}$,
$\omega_m\approx 10^4\,{\rm s}^{-1}$, 
$m\approx 10^{-14}\,{\rm Kg}$, 
$L\approx 1.5\,{\rm m}$, 
$Tr\approx 10^{-6}$ and $T\approx 10^{-7} K$.
Of course, other
choices satisfying the above mentioned criteria can be made giving 
the same qualitative results.
We are aware that a delicate point could be the realization of the 
mechanical oscillator with a
very small mass, but we would remark that the mass parameter could
 also be interpreted as an
effective value coming from the density of the vibrational modes 
of the mechanical oscillator
\cite{Heid}.
Furthermore, the discussed model could be improved by inserting 
an active Kerr medium
inside a cavity which enhances the nonlinear effects, slowing down 
the decoherence. 

Finally, even if we have not coupled the system under study with an 
external readout 
apparatus able to measure the momentum of the moving mirror, 
we think 
that the presented
model represents an interesting alternative way to approach, also in 
the experimental sense,
the quantum macroscopic coherence phenomena.

\section*{Acknowledgments}

This work has been partially supported by European 
Community under the 
Human Capital and Mobility (HCM) programme.
One of us, V. I. M. also gratefully acknowledges the 
University of 
Camerino for the kind
hospitality and the financial support of the Istituto 
Nazionale di Fisica Nucleare.

\newpage

FIGURE CAPTIONS

Fig. 1 The marginal distribution $P(X)$ 
is
plotted as function of the quadrature variable $X$ 
for $\kappa=0.5$, $|\alpha_0|=\sqrt{7}$, at $t^*=2\pi$ and 
in two different cases:
$\gamma=0$ (dashed line) and $\gamma=2\times 10^{-2}$ (solid line).

Fig. 2 The marginal distribution $P(X|y_{t'})$ 
is plotted as function of the quadrature variable $X$ 
for $y_{t'}=0$, 
$\kappa=0.52$,
$|\alpha_0|=\sqrt{7}$,
at $t'=3\pi/2$ and in three different cases:
$\gamma=0, N_{th}=0$ (dashed line); 
$\gamma=2\times 10^{-2}, N_{th}=0$ (solid line); 
$\gamma=2\times 10^{-2}, N_{th}\ge 20$ (dotted line).

Fig. 3 The correlation coefficient $C_{s,m}$ is plotted against the 
time for $\kappa=0.5$ in the case of  
$\gamma=0, N_{th}=0$ (dashed line); 
$\gamma=10^{-2}, N_{th}=0$ (solid line); 
$\gamma=10^{-2}, N_{th}=2$ (dotted line).

Fig. 4 The distribution ${\cal P}_{in}(n)$ vs. the photon number 
is plotted for a pseudo-cat with $\kappa=0.5$, 
$|\alpha_0|=\sqrt{7}$ and
$t=0.84\times2\pi$ in the case of
$\gamma=0, N_{th}=0$ a); 
$\gamma=10^{-2}, N_{th}=0$ b); 
$\gamma=10^{-2}, N_{th}=2$ c).

Fig. 5 The same of Fig. 4, but for ${\cal P}_{out}(n)$.

Fig. 6 The marginal distribution $P^{pc}(X)$ 
is plotted for 
$\kappa=0.5$, $|\alpha_0|=\sqrt{7}$,
$t=0.84\times 2\pi$ and the values of $\gamma$ and $N_{th}$ 
indicated in the figure.
It is also compared with the distribution for a pure cat.

\end{document}